\documentclass[times,twocolumn,final]{elsarticle}
\usepackage{medima}
\usepackage{framed,multirow}

\usepackage{amsmath,amssymb}
\usepackage{latexsym}
\usepackage{placeins}

\usepackage{url}
\usepackage{xcolor}
\usepackage{adjustbox}
\usepackage{threeparttable}
\usepackage{booktabs}
\usepackage{multirow}
\usepackage{tabularx}
\usepackage{ulem}

\usepackage{algorithm}
\usepackage{algorithmic}

\definecolor{airforceblue}{rgb}{0.36, 0.54, 0.66}
\definecolor{ballblue}{rgb}{0.004, 0.50, 0.69}
\definecolor{newcolor}{rgb}{.8,.349,.1}

\usepackage[colorlinks]{hyperref}
\usepackage[utf8]{inputenc}
\usepackage{graphicx}

\AtBeginDocument{%
  \hypersetup{
    citecolor=ballblue,
    linkcolor=ballblue,   
    urlcolor=ballblue}}

\journal{Medical Image Analysis}

\begin{document}

\verso{Jianwei Chen \textit{et~al.}}

\begin{frontmatter}
\title{Hierarchical Multiscale Structure-Function Coupling for Brain Connectome Integration\tnoteref{tnote1}}

\author[1]{Jianwei Chen}%\fnref{fn1}
\author[3]{Zhengyang Miao}%\fnref{fn1}
%\fntext[fn1]{Equal contribution.}
\author[5]{Wenjie Cai}
\author[4]{Jiaxue Tang}
\author[2]{Boxing Liu}
\author[1]{Yunfan Zhang}
\author[4]{Yuhang Yang}
\author[1]{Hao Tang}
\author[7]{Carola-Bibiane Schönlieb}
\author[6]{Zaixu Cui}
\author[4]{Du Lei}
\author[3]{Shouliang Qi\corref{cor1}}
\author[1,2,7,8]{Chao Li\corref{cor1}}

\cortext[cor1]{co-corresponding author.}
% \fntext[adni]{Data used in preparation of this article were obtained from the Alzheimer's Disease Neuroimaging Initiative (ADNI) database (adni.loni.usc.edu). As such, the investigators within the ADNI contributed to the design and implementation of ADNI and/or provided data but did not participate in analysis or writing of this report. A complete listing of ADNI investigators can be found at: http://adni.loni.usc.edu/wp-content/uploads/how_to_apply/ADNI_Acknowledgement_List.pdf}
\ead{cl647@cam.ac.uk}

\address[1]{School of Medicine, University of Dundee, UK}
\address[2]{School of Science and Engineering, University of Dundee, UK}
\address[3]{College of Medicine and Biological Information Engineering, Northeastern University, China}
\address[4]{College of Medical Informatics, Chongqing Medical University, China}
\address[5]{School of Airspace Science and Engineering, Shandong University, China}
\address[6]{Chinese Institute for Brain Research, China}
\address[7]{Department of Applied Mathematics and Theoretical Physics, University of Cambridge, UK}
\address[8]{Department of Clinical Neurosciences, University of Cambridge, UK}

% ...

\received{-}
\finalform{-}
\accepted{-}
\availableonline{-}
\communicated{-}

%%%
\begin{abstract}
Integrating structural and functional connectomes remains challenging because their relationship is non-linear and organized over nested modular hierarchies. We propose a hierarchical multiscale structure–function coupling framework for connectome integration that jointly learns individualized modular organization and hierarchical coupling across structural connectivity (SC) and functional connectivity (FC). The framework includes: (i) Prototype-based Modular Pooling (PMPool), which learns modality-specific multiscale communities by selecting prototypical ROIs and optimizing a differentiable modularity-inspired objective; (ii) an Attention-based Hierarchical Coupling Module (AHCM) that models both within-hierarchy and cross-hierarchy SC–FC interactions to produce enriched hierarchical coupling representations; and (iii) a Coupling-guided Clustering loss (CgC-Loss) that regularizes SC and FC community assignments with coupling signals, allowing cross-modal interactions to shape community alignment across hierarchies. We evaluate the model's performance across four cohorts for predicting brain age, cognitive score, and disease classification. Our model consistently outperforms baselines and other state-of-the-art approaches across three tasks. Ablation and sensitivity analyses verify the contributions of key components. Finally, the visualizations of learned coupling reveal interpretable differences, suggesting that the framework captures biologically meaningful structure–function relationships. The code is available at \href{https://github.com/chenminghe343-spec/Hierarchical-Multiscale-Structure-Function-Coupling-for-Brain-Connectome-Integration}{github}.
\end{abstract}

\begin{keyword}
%% MSC codes here, in the form: \MSC code \sep code
%% or \MSC[2008] code \sep code (2000 is the default)
% \MSC 41A05\sep 41A10\sep 65D05\sep 65D17
%% Keywords
\KWD Structure-function coupling  \sep Structural connectome \sep Functional connectome \sep Multi-modal graph learning
\end{keyword}

\end{frontmatter}

%\linenumbers

%% main text

\section{Introduction}

The brain connectome is a graphical representation of neural wiring in the brain, providing a unified framework for studying its complex, dynamic, hierarchical organization. Specifically, the brain is parcellated into regions using anatomical atlases, with structural connectivity (SC) among these regions inferred from white matter pathways via diffusion MRI (dMRI)~\citep{vskoch2022human}, and functional connectivity (FC) is estimated from co-activations of the BOLD signals from functional MRI (fMRI)~\citep{hu2024d}. These multimodal connectomes offer complementary insights into brain organization and its dynamics throughout development, aging~\citep{ball2014rich, chan2014decreased}, and various disorders~\citep{fornito2015connectomics,wei2021quantifying,wei2023predicting,sinha2020glioblastoma}. %wei23 21
%\textcolor{red}{Brain connectome examples}
Integrating multimodal connectomes can provide a comprehensive characterization of the brain. However, simple fusion strategies, such as concatenation~\citep{sebenius2021multimodal} or element-wise summation~\citep{zhang2021deep}, are limited in their ability to model non-linear interactions across multimodal connectomes and fail to reflect the complex brain organization. 

The structure-function coupling (SFC), as elucidated by neuroscience research, highlights how brain structure supports or constrains functional dynamics in physiological and pathological conditions~\citep{suarez2020linking,zhang2022predicting}. As such,  modeling SFC offers a promising approach to more comprehensively integrate multimodal connectomes with neuroscience interpretability. Recent studies have incorporated structure-function interactions explicitly as coupling signals to guide multimodal integration. For example, studies have proposed coupling SC and FC into a joint graph via learnable interaction edges, to predict brain age or disorders~\citep{li2022joint,xia2025interpretable}. Other studies learn SFC by modeling interactions between SC and FC representations through cross-modal alignment objectives such as mutual learning and contrastive learning~\citep{yang2023mapping,ye2023bidirectional}. Despite their success, these models typically operate at the regional level and tend to overlook the intrinsic organization of the brain.

Recent advances suggest that SFC follows a hierarchical, modular architecture~\citep{fotiadis2024structure,tang2025hierarchical}, where brain regions are organized into communities nested within coarser-grained modules across multiple levels of brain hierarchy, forming a multiscale partition of the brain connectome and %that spans coarse global modules to fine local communities 
 reflecting intrinsic brain organization~\citep{meunier2009hierarchical}. This observation has inspired connectome integration approaches that explore hierarchical SFC, such as network communication models~\citep{zamani2022local}, and the mapping of functional signals onto the structural graph harmonics~\citep{sun2024structure}. Despite advancements, these approaches often rely on handcrafted features and shallow aggregation strategies, which restrict their ability to capture complex, nonlinear interactions between modalities. 
 
More recent graph learning approaches explore SFC through regional subgraphs using graph traversal algorithms ~\citep{ye2023rh,huang2025local}, or community partitions based on prior functional modules~\citep{xia2024img,yeo2011organization}.
%相当于定义细粒度的是community，粗粒度是module.modular是性质，community是输出; modular organization = community structure
Despite efforts, these approaches face limitations. \textbf{First}, most methods rely on fixed modular templates or atlas-based parcellations, e.g., a canonical functional modular template~\citep{yeo2011organization}, before learning the SFC~\citep{xia2024img}. Another study~\citep{messe2020parcellation} derives SFC across resolutions using a family of cortical atlases that define increasing numbers of nodes, progressing from coarser to finer scales. However, rare studies account for individualized, modality-specific brain organization~\citep{fotiadis2024structure}, due to their reliance on pre-defined partitions from priors or atlases. \textbf{Second}, neuroscience research suggests that the brain is organized as a hierarchical system: within-hierarchy communities support specialized and segregated processing, while interactions across hierarchy are crucial for propagating information throughout the system~\citep{jiang2023information, pines2023development}.
However, existing models predominantly focus on within-hierarchy interactions, employing graph transformers~\citep{feng2025cross} or GNNs~\citep{xia2025interpretable}, and largely neglect cross-hierarchy interactions, leading to incomplete hierarchical coupling representations. \textbf{Third}, before modeling SFC, many studies derive structural and functional community structure using two separate community-detection methods~\citep{suarez2020linking,betzel2013multi,lurie2024cortical}. This approach may overlook cross-modal constraints in modeling the brain organization that jointly shape the community structure in SC and FC~\citep{seguin2022network}.

To address the above challenges, we propose a GNN-based framework that models hierarchical multiscale structure-function coupling (\textbf{HiM-SFC}) for integrating multimodal brain connectomes. \textbf{First}, we propose a Prototype-based Modular Pooling (PMPool) strategy with two modality-specific branches. Within each modality, PMPool identifies prototypical ROIs as community anchors and clusters them by prototype distances, yielding modality-specific community structure. To learn individualized community structures, a differentiable modularity-inspired pooling objective is introduced to encourage community assignments to approximate each connectome's modularity, rather than fixed partition. \textbf{Second}, we introduce an Attention-based Hierarchical Coupling Module (AHCM) to jointly model within- and cross-hierarchy structure–function interactions across scales. Specifically, within-hierarchy interactions are modeled via a bidirectional attention mechanism, while cross-hierarchy interactions are captured via a cross-attention mechanism and embedded as a coupling bias to enrich hierarchical coupling representations. \textbf{Lastly}, to incorporate cross-modal interactions into community structure learning, we introduce a Coupling-guided Clustering loss (CgC-Loss) that regularizes the SC and FC assignment matrices with coupling signals, enabling learned communities to be jointly constrained by both structure and function at each hierarchical level. Our main contributions are summarized as fourfold.
\begin{itemize}
    \item A GNN-based framework for integrating structural and functional connectomes through modelling hierarchical SFC, and jointly learns individualized, modality-specific hierarchical modular organization of the brain.

    \item A Prototype-based Modular Pooling (PMPool) strategy with two modality-specific branches that identify prototypical ROIs to assign communities, and a differentiable, modularity-inspired pooling objective encourages community structures to approximate individualized modularity.

    \item An Attention-based Hierarchical Coupling Module (AHCM) that captures both within- and cross-hierarchy structure–function interactions, employing bidirectional and cross attention mechanisms to aggregate multiscale interactions into rich hierarchical SFC representations.

    \item A Coupling-guided Clustering loss (CgC-Loss) that leverages coupling signals to regularize SC and FC assignment matrices at each hierarchical level, yielding community structure jointly constrained by structure and function.
\end{itemize}

\section{Related work}

\subsection{Hierarchical modular organization of the  connectome}

% 介绍脑连接组hierarchical modular组织是什么，结构和功能模块相关工作
Hierarchical modular organization refers to a nested community structure in which smaller communities are embedded within larger modules, forming a multiscale organization. In structural networks, this typically reflects locally dense clusters supported by short-range white matter connections, combined with long-range pathways for global integration~\citep{bullmore2009complex}. In functional networks, modular structure reflects coordinated activity patterns. A canonical example is the functional parcellations by Yeo \citep{yeo2011organization}, who derived 7- and 17-network templates by clustering resting-state fMRI connectivity patterns from a large population.

% 简要介绍深度GNN cluster pooling在研究hierarchical modular的贡献（如何diffpool），说明他们主要探究一般网络（论文引用、社交网络等），在脑网络组织结构探索上有限
% To learn hierarchical representations of brain connectomes, several studies have incorporated GNN pooling into end-to-end predictive frameworks. BrainGNN~\citep{li2021braingnn}, for example, performs ROI-selection pooling with dedicated regularization, enabling the model to highlight salient regions and expose community structure in fMRI while improving prediction and interpretability. To explicitly model the multiscale structure via clustering, \citet{zhang2024constructing} propose a Transformer-based framework with sparse attention that progressively coarsens functional brain networks into a hierarchy for early AD diagnosis. They additionally improve the starting scale by splitting atlas regions into finer sub-nodes. Despite these advances, existing methods examine hierarchical modular organization within each modality separately, largely overlooking how SC-FC interactions are constrained by brain organization and how these interactions could, in turn, shape community structure.

% 介绍hierarchical modular结构和SFC怎么相依相存，介绍几个用线性统计等方法的研究
SFC explicitly characterizes how anatomical wiring constrains functional interactions. Hierarchical modular organization and SFC are tightly linked, where coupling varies within and across hierarchical scales. Several studies have investigated this relationship using linear statistics and descriptive network analysis. \citet{zamani2022local} systematically benchmark a broad family of communication- and geometry-based predictors for SFC, showing that coupling is more informative and heterogeneous at the regional scale than at the whole-brain level. \citet{sun2024structure} quantify hierarchical structure-function discrepancies using a structural decoupling index derived from structural graph harmonics and demonstrate that Alzheimer’s disease exhibits a characteristic spatial pattern of abnormal decoupling that relates to cognition and supports diagnostics.

\subsection{Multimodal brain connectome integration}

% 介绍machine learning和CNN研究多模态脑连接组融合
Structural and functional connectomes provide complementary views of brain connectivity. Consequently, many studies have proposed methods to integrate both modalities. A common strategy is to concatenate modality-specific features and use a multilayer perceptron (MLP)~\citep{salas2010feature}. Convolutional approaches, such as BrainNetCNN \citep{kawahara2017brainnetcnn}, instead treat connectivity matrices as structured inputs and leverage operations over edge- and node-like representations to capture topological patterns. More recently, GNNs have become a prominent framework for multimodal integration because they can model non-Euclidean relationships in graph-structured data. For example, multi-modal GCNs (M-GCNs)~\citep{dsouza2021m} use subject-specific SC to guide learning on FC, whereas multi-view GCNs (MV-GCNs)~\citep{zhang2018multi} learn separate encoders for each modality and fuse the resulting representations for classification. Despite their progress, many of these approaches still behave largely as feature-level fusion schemes and have limited capacity to represent non-linear, region-specific cross-modal interactions.

% 介绍coupling概念和相关SFC深度方法，说明他们共同问题是将图当作一个flat graph建模coupling，忽略更复杂组织结构例如hierarchical modular

\subsection{Modeling SFC for connectome integration}
Modeling SFC, therefore, offers a principled route for multimodal integration, with the benefit of improved interpretability relative to naive feature combination. For instance, JointGCN \citep{li2022joint} models ROI-wise SFC by introducing learnable coupling edges between corresponding ROIs in SC and FC. Cross-GNN~\citep{yang2023mapping} further captures inter-modal dependencies via dynamic graph learning and mutual learning, and has been applied to disease classification. While these methods move beyond naive fusion, they typically model connectomes as flat graphs at a single scale (primarily at the ROI level), thereby ignoring the hierarchical modular structure of the brain. This risks overlooking key biological constraints on cross-modal interactions.

  \begin{figure*}[t]
    \centering
    \includegraphics[width=1\linewidth]{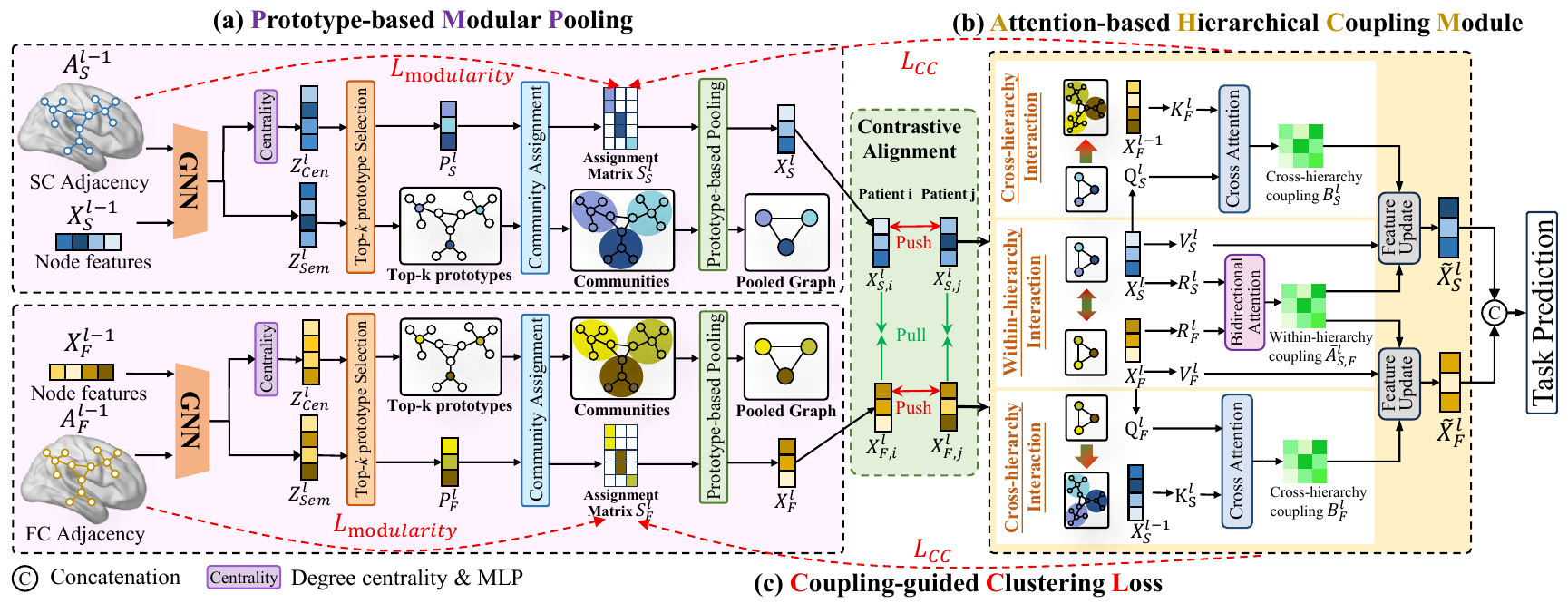}
    \caption{Framework of HiM-SFC. Paired SC and FC are encoded with a GNN and coarsened by (a) prototype-based modular pooling (PMPool) to obtain hierarchical community representations. (b) The attention-based hierarchical coupling module (AHCM) then models within- and cross-hierarchy structure–function interactions to produce coupling-enhanced features, while (c) a coupling-guided clustering loss (CgC-Loss) encourages cross-modal consistency of community assignments. The resulting hierarchical features are fused for the downstream prediction task.}
    \label{fig:framework}
\end{figure*}

\vspace{-.5em}
\subsection{Modeling Hierarchical modular SFC}

% 介绍几个使用深度学习建模hierarchical modular的方法，和intro类似说明他们存在的limitations，最后简要说明我们的研究目的，是为了解决什么问题或填补什么空缺

To capture non-linear SFC within a hierarchical modular structure, recent graph learning models have begun incorporating modular priors and multi-scale representations. IMG-GCN~\citep{xia2024img} uses canonical functional modules as guidance to compute module-specific structure-function interaction. \citet{feng2025cross} constrain FC self-attention using an SC-derived enhanced mask at each layer and apply cross-modal top-$k$ pooling to progressively coarsen the graph, enabling coupling and multiscale representations to be learned jointly for brain disease diagnosis. However, these methods primarily focus on within-level interactions (e.g., ROI-to-ROI or module-to-module), whereas cross-level coupling (e.g., ROI-to-module) is less explicitly represented.

Our aim is to address these gaps by learning hierarchical multiscale SFC in a data-driven and subject-specific manner, while jointly constraining the hierarchical organization of SC and FC, rather than learning them independently. In our framework, cross-modal interactions actively guide module formation and alignment across modalities, and the learned hierarchy, in turn, structures how coupling is modeled across multiple scales. This joint, multilevel design is intended to capture richer, more biologically plausible structure-function relationships, improve interpretability, and support stronger downstream predictions.

\vspace{-.5em}
\section{Methodology}

Fig.~\ref{fig:framework} demonstrates the overall framework. We treat paired SC and FC  as two graphs per subject, and input their adjacency matrices and node features. First, prototype-based modular pooling \textbf{(PMPool)} encodes each modality with a GNN, selects prototype ROIs, and learns an assignment matrix to pool ROIs into community-level super nodes, producing multiscale representations regularized by a modularity loss (Fig.~\ref{fig:framework}a). The attention-based hierarchical coupling block \textbf{(AHCM)} encourages SC and FC features to align via contrastive learning, then uses attention to model both within-hierarchy and cross-hierarchy structure function interactions, yielding hierarchical multiscale representations (Fig.~\ref{fig:framework}b). A coupling-guided clustering loss (\textbf{CgC-Loss}) uses the learned coupling strength to regularize SC and FC assignments, so cross-modal interactions can guide the assignment of the community (Fig.~\ref{fig:framework}c). Finally, the hierarchical features are fused for downstream prediction.

\subsection{Preliminary}

%The proposed framework is shown in Fig.~\ref{fig:framework}. 
Given a multimodal brain connectome dataset 
$\mathcal{D} = \{ \mathcal{P}_1, \dots, \mathcal{P}_P \}$ 
with $P$ subjects, each subject $\mathcal{P}_p$ is processed with a paired
structural connectome $G_S^{p}$ derived from dMRI, and a functional connectome
$G_F^{p}$ derived from rsfMRI. Each subject has a corresponding label $y^{p}$ for the prediction task.
Each graph $G = (V, A, X)$ consists of a node set $V$, an adjacency matrix
$A \in \mathbb{R}^{N \times N}$ and a node feature matrix
$X \in \mathbb{R}^{N \times d}$, where $N = |V|$ denotes the number of brain regions and $d$ is the feature dimension. In this study, the structural and functional node features $(X_S, X_F)$ are initialized by the corresponding
rows of the structural and functional adjacency matrices, respectively.
The goal of multimodal connectome fusion is to learn a mapping
$f : (G_S, G_F) \rightarrow y$ that predicts the label from paired structural and functional graphs.

\begin{figure}[!t]
  \centering
  \includegraphics[width=\columnwidth]{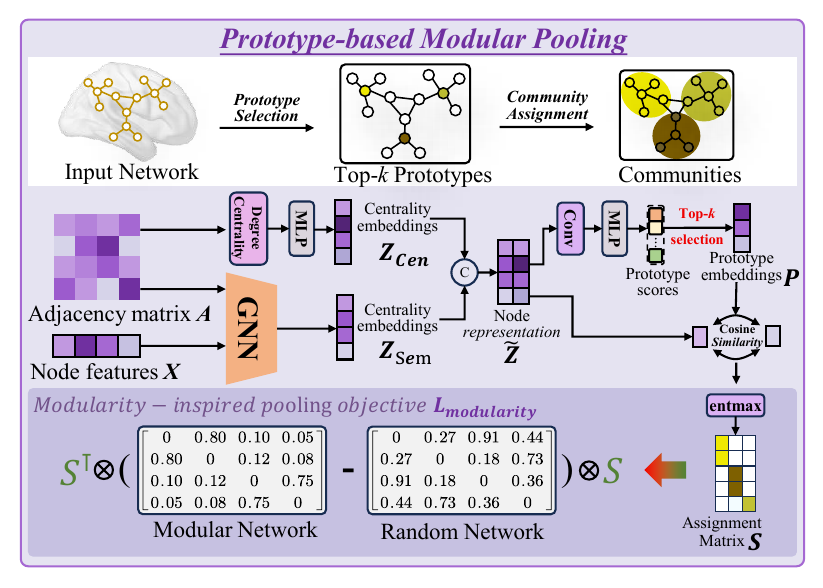}
  \caption{PMPool includes prototypical ROI selection and community assignment for pooling the connectome into a coarsened network. A modularity-inspired pooling objective encourages PMPool to capture the intrinsic modular structure of each subject’s connectome.}
  \label{fig:PMPool}
\end{figure}

\subsection{Prototype-based Modular Pooling}
\label{sec_PMPool}
\noindent\textbf{Prototype-based Pooling.}
Advanced studies have revealed that hierarchical structure-function interactions occur within the hierarchical modular organization. Therefore, our framework first aims to obtain the community structure for each modality. In this study, we develop a GNN pooling framework to simulate the hierarchical structure of the brain connectome. Each GNN pooling layer learns a community structure and guides the structure-function interactions. To learn how ROIs cluster within each hierarchy, we first identify the anchor of each cluster, namely the prototypical ROIs, and then learn how the remaining ROIs cluster around these prototypes. 

Consider a graph \(G=(V,A,X)\) for a given modality, where \(A\) is the adjacency matrix and \(X\) denotes the node features. For each node \(i\), we characterize its importance as a prototype based on both structural and semantic perspectives. To evaluate the structural importance, we compute the degree centrality for each node using 
\(A\), which reflects the extent to which a node is structurally connected to other nodes in the network, and embeds it as a centrality representation as follows:

\begin{equation}
    Z_{cen}=MLP\left( \frac{degree}{\max \left( degree \right)} \right). 
\end{equation}

In parallel, we use a one-layer GCN to embed node features \(X\) as semantic representations:

\begin{equation}
    Z_{sem}=GCN\left( A,\ X \right) =\sigma \left( \bar{A}XW \right), 
\end{equation}
where \(\bar{A}=D^{-\frac{1}{2}}AD^{-\frac{1}{2}}\) denotes the normalized adjacency matrix of \(A\), and \(D\) is the corresponding degree matrix with entries \(D_{ii}=\sum_j{A_{ij}}\), \(W\) is the learnable convolution weight matrix and \(\sigma\) is a non-linear activation function, for which we use the rectified linear unit (ReLU). Then the concatenated embeddings of these two representations \(\tilde{Z}=\left[ Z_{cen},\ Z_{sem} \right] \in \mathbb{R}^{N\times 2d'}\) are integrated as node representations. 

Subsequently, we convert the integrated representation \(\tilde{Z}\) into prototype scores that quantify the likelihood of each node acting as a community anchor. A convolution operation is employed with a filter \(h\in \mathbb{R}^{N\times 2d'}\) on the embeddings \(\tilde{Z}\), followed by a multilayer perceptron (MLP) to map the high-dimension representation for each node into a prototype score:

\begin{equation}
    U=MLP\left( h\,\,*\,\,\left( \tilde{Z} \right) \right),
\end{equation}
where \(U\in \mathbb{R}^{N\times 1}\), and a larger value of  \(U_i\) indicates that node \(i\) is more representative of its prototype. Next, we select top-$k$ prototypes using \(I_{\text{top-}k}=\operatorname{TopK}\left( U,k \right) \), and use the corresponding rows of \(\tilde{Z}\) to form the prototype embeddings \(P\in \mathbb{R}^{k\times 2d'}\). To determine community assignments, we compute the dot product similarity between all nodes and prototypes and apply the entmax function row-wise to obtain a sparse soft assignment matrix:

\begin{equation}
    S=entmax \left( \tilde{Z}P \right) \in \mathbb{R}^{N\times k}.
\end{equation}

The \(i^th\) row of the assignment matrix \(S\) indicates the sparse probability of node \(i\) being assigned to \(k\) communities.  Finally, a coarser graph \(G'=\left( V',\ A',\ X' \right) \) containing \(k\) super-nodes and aggregated features are obtained as follows:

\begin{equation}
    X'=S^{\top}X, \quad A'=S^{\top}AS.
\end{equation}

\noindent\textbf{Modularity-inspired Pooling Objective.}
Brain connectomes are known to exhibit modular organization, where regions form densely connected communities with comparatively sparse links between them. Modularity compares the observed connectivity within these communities by comparing it to what would be expected under a degree-preserving random graph, making it a standard measure for analyzing community structure. Building on this concept, we treat the soft assignment matrix \(S\) obtained above as a differentiable indicator for community membership following a previous method~\citep{tsitsulin2023graph}, and regularize it using an objective based on modularity:

\begin{equation}
    Q\left( A,\ S \right) =\frac{1}{2m}Tr\left( S^{\top}BS \right) ,
\end{equation}
where \(2m=\sum_i{d_i}\) denotes the total edge weight, and \(B=A-\frac{dd^{\top}}{2m}\). The \(Q\) is high when nodes that share strong assignments in \(S\) are more densely connected than expected from their degrees. We then follow the deep modularity formulation in graph clustering and derive the final optimized objective:

\begin{equation}
    L_{modularity}=-Q\left( A,\ S \right) +\frac{\sqrt{k}}{N}\lVert \sum_i{S_{i}^{\top}} \rVert _F-1,
\end{equation}
where \(\frac{\sqrt{k}}{N}\lVert \sum_i{S_{i}^{\top}} \rVert _F-1\) is a collapse regularizer that penalizes degenerate solutions where all nodes fall into a single community. By optimizing the loss function \(L_{modularity}\), the assignment matrix \(S\) is encouraged to group densely connected ROIs into the same community, guiding the PMPool to capture each subject's intrinsic community structure rather than forming arbitrary clusters. By stacking several PMPooI layers, we obtain an individualized hierarchical multiscale representation as the basis for the subsequent hierarchical SFC module.

\begin{figure}[!t]
  \centering
  \includegraphics[width=\columnwidth]{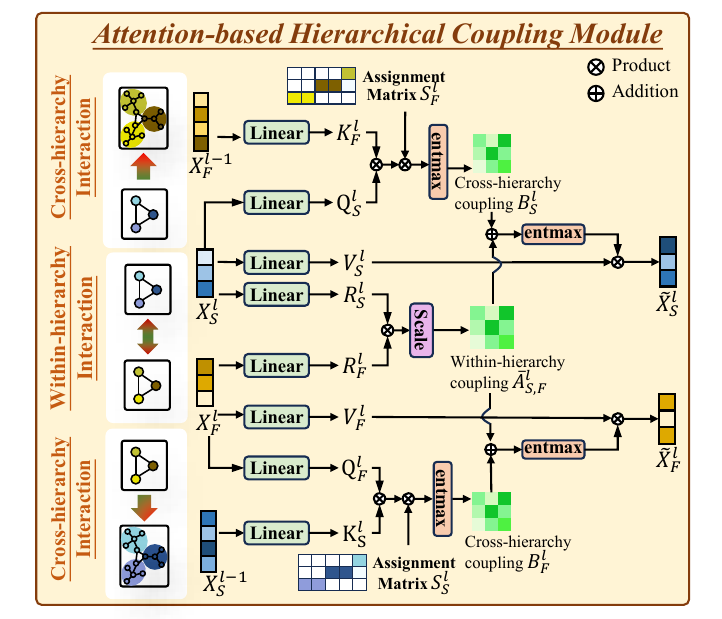}
  \caption{Architecture of AHCM. Within-hierarchy coupling is computed via bidirectional attention, while cross-hierarchy coupling is captured by two cross-attention branches. These hierarchical coupling signals are aggregated to obtain enriched hierarchical multiscale representations.}
  \label{fig:LCC}
\end{figure}

\subsection{Attention-based Hierarchical Coupling Module}
\label{sec_AHCM}
After obtaining the pooled embeddings from PMPool, it remains challenging to obtain fine-grained coupling from the derived attention weights, as structural and functional representations live in separate feature spaces. To address this, we first implement a lightweight contrastive alignment between the two modalities at each hierarchical level. Specifically, given the structural and functional community embeddings \(X_{S}^{l}\) and \(X_{F}^{l}\), we generate graph-level summaries using mean pooling and then process them through small projection heads. We apply an InfoNCE-style loss to bring together representations from the same subject across modalities while pushing apart representations from different subjects. This process aligns \(X_{S}^{l}\) and \(X_{F}^{l}\) into a latent space, allowing subsequent attention to reflect the coupling between structure and function rather than merely trivial differences in scale or orientation.

\noindent\textbf{Within-hierarchy Interaction.} Rather than employing two independent cross-modal attention blocks, we model the SFC at layer \(l\) using a specifically designed bidirectional attention mechanism. We first apply linear projections to the embeddings of both modalities and then compute a similarity matrix \(\bar{A}_{S,F}^{l}\), which is shared between the structural and functional representations, as follows:
\begin{equation}
    \bar{A}_{S,F}^{l}=\frac{R_SR_{F}^{\top}}{\sqrt{d}}\in \mathbb{R}^{N_l\times N_l},
\end{equation}
where \(\ R_{S}=X_{S}^{l}W_R,\ R_{F}=X_{F}^{l}W_R\). This shared-similarity matrix captures bidirectional coupling symmetrically.

\noindent\textbf{Cross-Hierarchy Interaction.} Interactions within a single hierarchy are insufficient for capturing couplings across different hierarchical levels, which are essential for integrating and propagating information throughout the system. To address this, we introduce the concept of cross-hierarchy interaction and utilize it to create a bias term that adjusts the shared similarity matrix. For example, in the case of the functional connectome, let \(X_{S}^{l-1} \in \mathbb{R}^{N_{l-1} \times d}\) represent the finer structural regional embeddings at level \(l-1\), and let \(S_{S}^{l} \in \mathbb{R}^{N_{l-1} \times N_l}\) denote the structural assignment matrix that connects regions to communities. We begin by computing attention from functional communities at level \(l\) to structural regions at level \(l-1\) through cross-hierarchy attention:

\begin{equation}
H_{F\rightarrow S}^{l}=entmax\left( \frac{Q_FK_{S}^{\top}}{\sqrt{d}} \right) \in \mathbb{R}^{N_l\times N_{l-1}},
    % H_{F\rightarrow S}^{l}=entmax\left( \frac{Q_FK_{S}^{\top}}{\sqrt{d}} \right) ,\quad Q_F=W_QX_{F}^{l}\in \mathbb{R}^{N_l\times d},K_S=W_KX_{S}^{l}\in \mathbb{R}^{N_{l-1}\times d}
\end{equation}
where \(Q_F=W_QX_{F}^{l},K_S=W_KX_{S}^{l-1}\). \(H_{F\rightarrow S}^{l}\) represent functional communities and their relationships with structural regions. 

\noindent\textbf{Hierarchical Interactions Aggregation.} We then aggregate the relationships using the structural assignments with \(B_{F}^{l}=\left(H_{F\rightarrow S}^{l}S_{S}^{l}\right)^{\top}\in \mathbb{R}^{N_l\times N_l}\). \(B_{F}^{l}\) can be viewed as a coupling bias, which incorporates information about how cross-hierarchy support for coupling between functional communities and structural regions. Finally, this bias matrix is aggregated to the shared similarity matrix \(\bar{A}_{S,F}^{l}\). To obtain the hierarchy coupling from the functional to structural viewpoint, we apply the column-wise entmax function, which updates the functional representations as follows: 

\begin{equation}
    \tilde{X}_{F}^{l}=entmax_{col}\left( \bar{A}_{S,F}^{l}+ B_{F}^{l}\right) V_F\in \mathbb{R}^{N_l\times d},
\end{equation}
where \(V_F=W_VX_{F}^{l}\). The hierarchy coupling from the structure to function view, and the updated structural representations can be computed in a symmetric manner. In this way, we obtain enriched hierarchical multiscale representations \(\tilde{X}_{F}^{l}\) and \(\tilde{X}_{S}^{l}\) for downstream prediction tasks.

\begin{figure}[!t]
  \centering
  \includegraphics[width=\columnwidth]{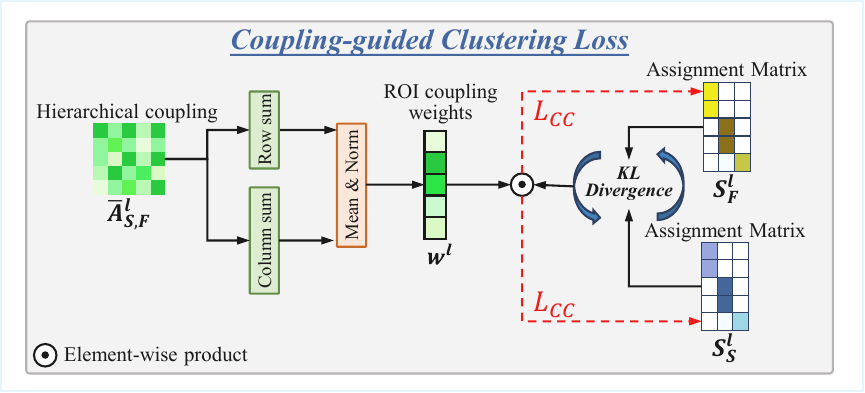}
  \caption{The coupling-guided clustering loss $L_{CC}$. ROI-wise coupling weights are computed from the hierarchical coupling matrix via row and column aggregation and normalization, and used to weight a symmetric KL divergence between SC and FC assignment matrices, promoting cross-modal consistency driven by SFC.}
  \label{fig:LCC}
\end{figure}

\subsection{Coupling-guided Clustering Loss Function}
Prior knowledge indicates that the hierarchical modular structure within the brain connectome is not arbitrary but organized according to the pattern of structure-function coupling\citep{seguin2022network,betzel2013multi}. In our framework, PMPool is first used to learn hierarchical modular organizations from structural and functional connectomes, which are denoted by the structural and functional assignment matrices at each level. Next, we convert the coupling at each hierarchy into meaningful signals to guide the community assignments for both modalities. With the shared similarity matrix \(\bar{A}_{S,F}^{l}\) at level \(l\), we first compute a region-wise coupling weight from bidirectional coupling views through the following process:

\begin{equation}
    w_{i}^{l}=normalize\left( \frac{1}{2}\left( \sum_j{|\bar{A}_{S,F}^{l}\left( i,j \right) |+\sum_j{|\bar{A}_{S,F}^{l}\left( j,i \right) |}} \right) \right).
\end{equation}

Intuitively, ROIs with stronger coupling weights exhibit more similarity in their structural and functional assignment matrices. Therefore, we can view \(i^{th}\) rows of the assignment matrices \(s_{S,i}^l\) and \(s_{F,i}^l\) as probability vectors and calculate the distribution distance between them. Then we use the coupling weights to modulate how similar they are:

\begin{equation}
    L_{CC}=\frac{1}{N_l}\sum_{i=1}^{N_l}{\frac{w_{i}^{l}}{2}}\left( KL\left( s_{S,i}^{l}||s_{F,i}^{l} \right) + KL\left( s_{F,i}^{l}||s_{S,i}^{l}\right) \right),
\end{equation}
where \(KL\left( \cdot \right) \) denotes the Kullback-Leibler divergence that is used to measure the distance between distributions. Together with the PMPool, this yields an individualized modular structure that is not only internally coherent within each modality but also maintains cross-modality consistency in regions of the brain where structure and function are closely connected.

\begin{algorithm}[!t]
\caption{Two-stage training procedure of HiM-SFC}
\small
\label{alg:framework_HiM-SFC}
\begin{algorithmic}[1]

\REQUIRE Paired structural and functional connectomes with labels $(G_S, G_F, y)$
\ENSURE Prediction $\hat{y}$

\STATE Initialize parameters of PMPool, AHCM, and the prediction MLP

\STATE \textbf{Stage 1: Self-supervised optimization of PMPool}
\FOR{epoch $=1$ to $E_1$}
    \STATE Extract node representations $\tilde{Z}_S$ and $\tilde{Z}_F$ from $G_S$ and $G_F$
    \STATE Select top-$k$ prototype nodes based on $\tilde{Z}_S$ and $\tilde{Z}_F$
    \STATE Compute assignment matrices $S_S$ and $S_F$
    \STATE Pool connectomes hierarchically: $X' = S^\top X$, $A' = S^\top A S$
    \STATE Compute modularity-inspired pooling loss $L_{\mathrm{modularity}}$
    \STATE Update PMPool parameters by minimizing $L_{\mathrm{modularity}}$
\ENDFOR

\STATE \textbf{Stage 2: Supervised optimization of the full framework}
\FOR{epoch $=1$ to $E_2$}
    \STATE Obtain multiscale SC and FC representations using pretrained PMPool
    \STATE Model hierarchical structure--function coupling with AHCM and obtain enriched representations $\tilde{X}_S$ and $\tilde{X}_F$
    \STATE Concatenate hierarchical representations to predict $\hat{y}$
    \STATE Compute task loss $L_{\mathrm{task}}$ and CgC-Loss $L_{CC}$
    \STATE Compute total loss $L_{\mathrm{total}} = L_{\mathrm{task}} + \lambda L_{CC}$
    \STATE Update model parameters by minimizing $L_{\mathrm{total}}$
\ENDFOR

\RETURN $\hat{y}$

\end{algorithmic}
\end{algorithm}

\subsection{Training Objective and Optimization}

To effectively capture hierarchical multiscale SFC, we adopt a two-stage training strategy. Because hierarchical coupling relies on community structure, we optimize the modularity-inspired pooling objective \(L_{modularity}\) of the PMPool module in a self-supervised manner in stage one, so as to obtain stable and individualized community assignments for structural and functional connectomes. After establishing this modality-specific modular organization, we introduce the AHCM and jointly optimize the full framework in a supervised manner in stage two. The enriched hierarchical multiscale representations \(\tilde{X}_{F}\) and \(\tilde{X}_{S}\) produced by AHCM are concatenated and passed to an MLP for task-specific prediction. Meanwhile, the Coupling-guided Clustering loss \(L_{CC}\) is incorporated as a regularization term during supervised training, where it leverages learned hierarchical coupling signals to constrain and refine the structural and functional community assignment matrices. The detailed training procedure of HiM-SFC can be found in Algorithm~\ref{alg:framework_HiM-SFC}.

% --------------------------------------------

%\input{algorithms/algorithm1}

% \begin{table}[!t]

\section{Implementation details}

% \begin{table}[!t]
% \centering
% \setlength{\tabcolsep}{4pt}
% \fontsize{8}{10}\selectfont
% \renewcommand{\arraystretch}{0.95}
% \caption{Subject information on the datasets used in this study.}
% \label{Tab:DatasetInfo}
% \begin{tabular}{lccccc}
% \toprule
% \multirow{2}{*}{\textbf{Data set}} &
% \multicolumn{2}{c}{\textbf{Gender (M/F)}} &
% \multicolumn{2}{c}{\textbf{Age (Years)}} &
% \multirow{2}{*}{\textbf{Total}} \\
% \cmidrule(lr){2-3}\cmidrule(lr){4-5}
% & \textbf{Control} & \textbf{Patient} & \textbf{Control} & \textbf{Patient} & \\
% \midrule
% HCPYA  & 113/128  & — & 28.17 $\pm$ 3.67 & — & 241 \\
% NKI-Rockland  & 114/82  & —   & 34.96 $\pm$ 19.99 & — & 196 \\
% UKBiobank  & 1500/1500   & —   & 55.33 $\pm$ 7.66 & — & 3000 \\
% ADNI & 27/45 & 37/30 & 69.26 $\pm$ 8.60 & 72.48 $\pm$ 7.29 & 139 \\
% \bottomrule
% \end{tabular}
% \end{table}

\begin{table*}[!t]
\centering
\setlength{\tabcolsep}{3pt}
\fontsize{8}{10}\selectfont
\renewcommand{\arraystretch}{0.9}
\caption{Comparison results on HCPYA, NKI-Rockland and UKB for brain age prediction based on FC and SC.}
\label{Tab:CompareAge}

\begin{tabular*}{\textwidth}{@{\extracolsep{\fill}}lcc cc cc}
\toprule
\multirow{2}{*}{\textbf{Method}} &
\multicolumn{2}{c}{\textbf{HCPYA}} &
\multicolumn{2}{c}{\textbf{NKI-Rockland}} &
\multicolumn{2}{c}{\textbf{UKB}} \\
\cmidrule(lr){2-3}\cmidrule(lr){4-5}\cmidrule(lr){6-7}
& \textbf{MAE~$\downarrow$} & \textbf{RMSE~$\downarrow$}
& \textbf{MAE~$\downarrow$} & \textbf{RMSE~$\downarrow$}
& \textbf{MAE~$\downarrow$} & \textbf{RMSE~$\downarrow$} \\
\midrule
GCN~\citep{kipf2016semi}       & 4.46 (0.62) & 6.02 (1.26)
                               & 16.20 (1.49) & 21.67 (2.78)
                               & 5.97 (0.23) & 7.15 (0.29)\\
GIN~\citep{xu2018powerful}     & 4.17 (0.60) & 5.16 (0.55)
                               & 17.70 (2.44) & 23.40 (3.43)
                               & 6.17 (0.67) & 7.39 (0.78) \\
GraphSage~\citep{hamilton2017inductive}  
                               & 4.15 (0.43) & 5.50 (0.83)
                               & 15.96 (2.08) & 18.48 (1.31)
                               & 6.06 (0.51) & 8.58 (2.87) \\
M-GCN~\citep{dsouza2021m}      & 3.45 (0.34) & 4.36 (0.52)
                               & 17.36 (0.49) & 20.16 (0.56)
                               & 7.59 (0.57) & 12.55 (0.47) \\
MME-GCN~\citep{liu2022enhanced}
                               & 6.02 (1.38) & 7.14 (1.32) 
                               & 14.41 (0.34) & 19.44 (0.52) 
                               & 6.64 (0.65) & 7.93 (0.72) \\
Joint-GCN~\citep{li2022joint}  & 3.34 (0.41) & 4.15 (0.43)
                               & 13.70 (0.57) & 17.91 (0.76)
                               & \uline{5.47 (0.12)} & \uline{6.71 (0.07)}\\
Cross-GNN~\citep{yang2023mapping}   
                               & 5.21 (0.33) & 6.32 (0.35)
                               & \uline{13.25 (1.76)} & \textbf{16.62 (1.85)}
                               & 6.82 (0.10) & 7.56 (0.11) \\
MS-Inter-GCN~\citep{xia2025interpretable}
                               & 3.33 (0.36) & 3.95 (0.50) 
                               & 15.76 (1.07) & 19.04 (1.50)  
                               & 5.74 (0.17) & 6.85 (0.10) \\
RH-BrainFS~\citep{ye2023rh}
                               & \uline{3.28 (0.27)} & \uline{3.93 (0.31)} 
                               & 13.42 (2.46) & 17.84 (2.21)    
                               & 5.95 (0.53) & 7.29 (0.49) \\
IMG-GCN~\citep{xia2024img}     & 3.97 (0.36) & 4.59 (0.38)
                               & 14.04 (1.49) & 17.63 (1.31)
                               & 5.57 (0.46) & 6.74 (0.50) \\
CBGT~\citep{feng2025cross}     & 3.46 (0.15) & 4.18 (0.11)
                               & 14.63 (0.76) & 19.24 (0.32)
                               & 5.89 (0.15) & 7.68 (0.16) \\
\midrule
\textbf{HiM-SFC(ours)}           & \textbf{2.77 (0.17)} & \textbf{3.42 (0.19)}
                               & \textbf{12.86 (0.69)} & \uline{16.92 (0.59)}
                               & \textbf{5.35 (0.08)} & \textbf{6.50 (0.08)} \\
\bottomrule
\end{tabular*}
\end{table*}

\begin{table*}[!t]
\centering
\setlength{\tabcolsep}{3pt}
\fontsize{8}{10}\selectfont
\renewcommand{\arraystretch}{0.9}
\caption{Comparison results on HCPYA and UKB for MMSE and fluid intelligence score prediction based on FC and SC.}
\label{Tab:CompareCognitive}

\begin{tabular*}{\textwidth}{@{\extracolsep{\fill}}lcc cc}
\toprule
\multirow{2}{*}{\textbf{Method}} &
\multicolumn{2}{c}{\textbf{HCPYA (MMSE)}} &
\multicolumn{2}{c}{\textbf{UKB (Fluid Intelligence Score)}} \\
\cmidrule(lr){2-3}\cmidrule(lr){4-5}
& \textbf{MAE~$\downarrow$} & \textbf{RMSE~$\downarrow$}
& \textbf{MAE~$\downarrow$} & \textbf{RMSE~$\downarrow$}\\
\midrule
GCN~\citep{kipf2016semi}        & 3.73 (0.57) & 5.18 (1.43)
                                & 3.17 (0.26) & 3.69 (0.23)\\
GIN~\citep{xu2018powerful}      & 3.52 (0.43) & 4.09 (0.30)
                                & 3.60 (0.30) & 4.01 (0.43)\\
GraphSage~\citep{hamilton2017inductive}  
                                & 3.65 (0.68) & 5.18 (1.43)
                                & 3.14 (0.43) & 3.66 (0.40)\\
M-GCN~\citep{dsouza2021m}       & \uline{1.48 (0.11)} & \uline{1.95 (0.21)}
                                & 3.23 (1.43) & 5.30 (3.98) \\
MME-GCN~\citep{liu2022enhanced}
                               & 6.12 (0.33) & 6.61 (0.32)  
                               & 2.72 (0.06) & 3.17 (0.09) \\
Joint-GCN~\citep{li2022joint}   & 3.95 (0.95) & 4.61 (0.96)
                                & 2.18 (0.31) & 2.69 (0.33) \\
Cross-GNN~\citep{yang2023mapping}   
                                & 5.59 (0.44) & 5.68 (0.44)
                                & 2.88 (0.18) & 3.39 (0.18)\\
MS-Inter-GCN~\citep{xia2025interpretable}
                               & 1.60 (0.04) & 2.02 (0.18)
                               & \uline{1.91 (0.04)} & 2.20 (0.05) \\
RH-BrainFS~\citep{ye2023rh}11
                               & 1.64 (0.11) & 2.17 (0.23)   
                               & 1.93 (0.04) & 2.22 (0.06) \\
IMG-GCN~\citep{xia2024img}      & 1.72 (0.05) & 2.13 (0.03)
                                & 2.04 (0.05) & \uline{2.18 (0.06)} \\
CBGT~\citep{feng2025cross}      & 1.78 (0.19) & 2.06 (0.22)
                                & 2.10 (0.15) & 2.59 (0.17) \\
\midrule
\textbf{HiM-SFC(ours)}            & \textbf{1.01 (0.24)} & \textbf{1.25 (0.25)}
                                & \textbf{1.66 (0.04)} & \textbf{2.07 (0.05)} \\
\bottomrule
\end{tabular*}
\end{table*}

\begin{table*}[!t]
\centering
\setlength{\tabcolsep}{4pt}
\fontsize{8}{10}\selectfont
\renewcommand{\arraystretch}{0.9}
\caption{Comparison results on ADNI for disease classification based on FC and SC.}
\label{Tab:CompareDisease}
\begin{tabular*}{\textwidth}{@{\extracolsep{\fill}}lcccc}
\toprule
\textbf{Method} & 
\textbf{ACC}~$\uparrow$ & 
\textbf{AUC}~$\uparrow$ & 
\textbf{Sen}~$\uparrow$ & 
\textbf{Spec}~$\uparrow$ \\
\midrule
GCN~\citep{kipf2016semi}       
& 0.68 (0.06) & 0.60 (0.14) & 0.50 (0.15) & 0.74 (0.16) \\

GIN~\citep{xu2018powerful}        
& 0.64 (0.04) & 0.71 (0.03) & 0.61 (0.16) & 0.68 (0.15) \\

GraphSAGE~\citep{hamilton2017inductive}  
& 0.68 (0.07) & 0.62 (0.14) & 0.67 (0.15) & 0.69 (0.15) \\

M-GCN~\citep{dsouza2021m}
& 0.73 (0.07) & 0.70 (0.06) & 0.61 (0.23) & 0.74 (0.22) \\

MME-GCN~\citep{liu2022enhanced}
& 0.74 (0.07) & 0.72 (0.10) & 0.66 (0.14) & 0.79 (0.08) \\

Joint-GCN~\citep{li2022joint}
& 0.75 (0.08) & 0.70 (0.08) & 0.61 (0.20) & 0.79 (0.21) \\

Cross-GNN~\citep{yang2023mapping}
& 0.70 (0.10) & 0.71 (0.12) & 0.51 (0.25) & 0.73 (0.24) \\

MS-Inter-GCN~\citep{xia2025interpretable}
& 0.72 (0.05) & 0.75 (0.12) & 0.65 (0.21) & 0.71 (0.18) \\

RH-BrainFS~\citep{ye2023rh}
& 0.73 (0.14) & 0.75 (0.05) & 0.62 (0.26) & 0.75 (0.33) \\

IMG-GCN~\citep{xia2024img}
& 0.75 (0.08) & \uline{0.77 (0.08)} & 0.67 (0.15) & 0.76 (0.13) \\

CBGT~\citep{feng2025cross}
& \uline{0.76 (0.05)} & 0.76 (0.03) & \uline{0.68 (0.13)} & \uline{0.79 (0.13)} \\

\midrule
\textbf{HiM-SFC(ours)}       
& \textbf{0.78 (0.07)} & \textbf{0.79 (0.10)} & \textbf{0.69 (0.10)} & \textbf{0.81 (0.13)} \\
\bottomrule
\end{tabular*}
\end{table*}

\subsection{Datasets}
We evaluate our model on brain age and cognition prediction, as well as on disease classification. For brain-age prediction, we utilized three healthy cohorts. The first includes 196 subjects aged 4-85 years from the Nathan Kline Institute Rockland Sample (NKI-Rockland)~\citep{nooner2012nki}, with corresponding T1w, rsfMRI, and DTI images, preprocessed using MRtrix3 toolbox (\href{https://www.mrtrix.org/}{https://www.mrtrix.org/})~\citep{tournier2019mrtrix3}. SC and FC were constructed with 188 ROIs, based on the Craddock 200 spectral clustering atlas~\citep{craddock2012whole} in the prior study \citep{brown2012ucla}. The second includes 241 subjects aged 22-35 years from the Human Connectome Project Young Adult (HCPYA) dataset (release S900)~\citep{van2013wu}. SC and FC were constructed from minimally preprocessed T1w, rsfMRI, and diffusion MRI using the MRtrix3~\citep{chen2024group} and eXtensible Connectivity Pipelines (XCP-D; \href{https://xcp-d.readthedocs.io/en/latest/}{https://xcp-d.readthedocs.io/en/latest/})~\citep{mehta2024xcp}, respectively, following the Schaefer atlas \citep{schaefer2018local} containing 400 ROIs. 
Due to limited computational resources, the third includes 3,000 subjects aged 40-69 years from the UKBiobank (UKB)~\citep{alfaro2018image}, selected via stratified sampling from the $\sim$40,000 publicly released imaging participants as a subset. The MRI images were preprocessed using the UKB standard pipeline. SC and FC were constructed from the prior study~\citep{di2023connectomes} using the Schaefer atlas, which contains 500 ROIs \citep{di2023connectomes}. For cognitive score prediction, we use the same cohort from HCPYA and UKB, with Mini-Mental State Examination (MMSE) scores for HCPYA (range 23-30) and Fluid Intelligence Score for the UKB (range 0-13), respectively. For disease classification, we curated a subset of the Alzheimer’s Disease Neuroimaging Initiative (ADNI)~\citep{petersen2010alzheimer} cohort by selecting baseline ADNI3 participants with high-resolution multi-shell diffusion MRI and multiband resting-state fMRI, ensuring sufficient data quality for reliable structural and functional connectome construction. This resulted in 72 healthy controls and 67 patients with dementia, aged 51-90 years. We reproduce the UKB standard pipeline on preprocessed multimodal MRI and construct the SC and FC, following the Schaefer atlas \citep{schaefer2018local} with 500 ROIs. %Further detailed information on the datasets is shown in Table~\ref {Tab:DatasetInfo}.

% Preamble:
% \usepackage{booktabs}
% \usepackage{multirow}

For each dataset, the structural edges represent the streamline counts between ROIs, while functional edges represent the Pearson correlation coefficients between ROIs. This study focuses on positive functional connectivity and retains the top 20\% of strongest connections in each FC matrix to create a sparse functional network \citep{van2010exploring}. %More details about each datasets are shown in Table \ref{tab:datasets_details}.

% \begin{figure*}[!t]
%     \centering
%     \includegraphics[width=1\linewidth]{pdf/HCP_MMSE_flatcouple.pdf}
%     \caption{Visualization of learned SFC by AHCM from two different MMSE cognitive score groups (28 and 30) on HCPYA dataset.}
%     \label{fig:HCP_MMSE_flatcouple}
% \end{figure*}

\subsection{Implementation details}
The experiments of this study were conducted using PyTorch and Pytorch Geometric on an NVIDIA RTX A5000 GPU with 24GB of memory. We employed 5-fold cross-validation and reported the results in \(mean~(std)\). All hyperparameters were fine-tuned to achieve optimal performance on each dataset. The number of pooling ratio \(r\) for SC and FC in PMPool was set to 0.12. The regularization weight of \(L_{CC}\) was established at 0.11. As will be studied later, increasing GNN and pooling layers can make the model learn smoother representations, which impairs its expressive power on prediction tasks. Therefore, we set the number of pooling layers and GNN encoders as 1 and 2, respectively. The proposed framework was trained for 300 epochs, with a batch size of 32 and a learning rate of 0.001 using the Adam optimizer \citep{adam2014method}.

\begin{figure*}[!t]
  \centering
  \includegraphics[width=0.65\textwidth]{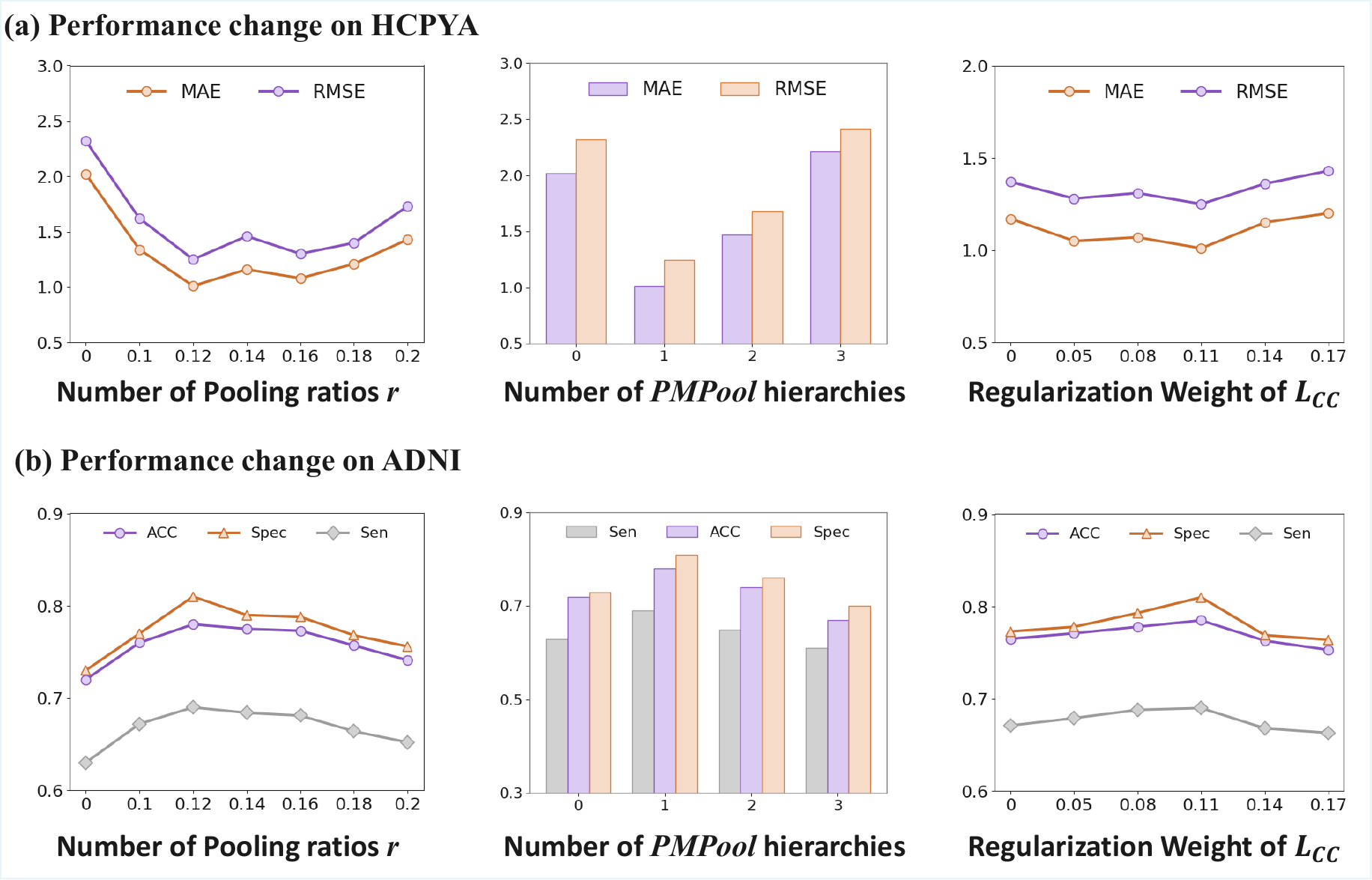}
  \caption{Cognitive score prediction performance of the proposed method with different pooling ratios, number of hierarchies, and regularization weight of \(L_{CC}\) on (a) HCPYA and (b) ADNI.}
  \label{fig:hyperparameter}
\end{figure*}

\subsection{Comparison models}

%\noindent\textbf{Brain age and cognitive score prediction.}
We evaluate the effectiveness of our proposed method by comparing it with various competing approaches across multiple datasets. The competing methods are grouped into three categories: (i) multimodal GNN baselines with simple fusion (e.g., feature concatenation), (ii) state-of-the-art SFC-based methods that model SFC on a flat graph, and (iii) recent SFC-based approaches that further incorporate hierarchical or modular organization of the connectome. \textbf{(i) Multimodal GNN baselines.} We include three baselines, GIN \citep{xu2018powerful}, GCN \citep{kipf2016semi} and GraphSAGE \citep{hamilton2017inductive}, as well as earlier GNN-based connectome fusion methods that use simple feature fusion, i.e., M-GCN \citep{dsouza2021m} and MME-GCN~\citep{liu2022enhanced}, which do not explicitly model non-linear structure–function interactions. \textbf{(ii) Flat-graph SFC SOTAs.} We include four coupling-based methods that explicitly model the non-linear structure-function interaction.  Among these, Joint-GCN \citep{li2022joint} is one of the first approaches to model SFC using GNNs, leveraging cross-modal interactions for prediction. Similarly, Cross-GNN \citep{yang2023mapping}  and the more recent MS-Inter-GCN~\citep{xia2025interpretable} propose learning multimodal interactions between SC and FC, which typically capture coupling from a flat graph perspective, ignoring the intricate brain connectome organization. \textbf{(iii) Hierarchical/modular SFC SOTAs.} CBGT \citep{feng2025cross} and IMG-GCN \citep{xia2024img} model SFC in a hierarchical or modular manner. RH-BrainFS~\citep{ye2023rh} proposes to learn indirect SFC through sampled local subgraph structure. Yet, they do not fully characterize individualized hierarchical modular organization, or how such an organization both constrains and is shaped by SFC. Table~\ref{Tab:CompareAge},~\ref{Tab:CompareCognitive} and~\ref{Tab:CompareDisease} demonstrate the comparison results on HCPYA, NKI-Rockland, UKB and ADNI datasets.

\subsection{Comparison metrics}
\label{section_comparison} 

 We report performance using mean absolute error (MAE), root mean squared error (RMSE) for both brain-age and cognitive-score prediction; accuracy (ACC), sensitivity (Sen), specificity (Spec) and the area under the receiver operating characteristic curve (AUC) for disease classification.

% \begin{figure}[!t]
%   \centering
%   \includegraphics[width=\columnwidth]{pdf/AB_HCPYA_ANDI.pdf}
%   \caption{Cognitive score prediction performance of the proposed method with different numbers of (a) pooling ratios, (b) hierarchy and (c) regularization weight of \(L_{CC}\)}
%   \label{fig:hyperparameter}
% \end{figure}

\section{Results \& Discussion}
\subsection{Comparison studies}
\subsubsection{Brain age prediction}
%\noindent\textbf{Brain age prediction.} 
To assess the effectiveness and generalization of the proposed model, we conduct brain age prediction on three datasets that span a wide age range (HCPYA focuses on young adults (21-35 yrs), UKB targets mid-to-late adults (40-70 yrs), NKI-Rockland includes healthy participants across a broad range (4-85 yrs). Results for all eight competing methods on brain age prediction are demonstrated in Table.~\ref{Tab:CompareAge}. The three multimodal GNN baselines achieve the worst performance across the three datasets, primarily due to their inability to capture the SFC associated with aging. By contrast, the models with explicit structure-function coupling, e.g., Joint-GCN and Cross-GNN, achieve superior performance compared to simple multimodal GNN fusion methods. In particular, the Joint-GCN improves performance to 3.34 MAE on HCPYA and 5.47 MAE on UKB, probably attributed to its ROI-to-ROI coupling design. Other state-of-the-art methods, i.e., IMG-GCN and CBGT, also outperform the GNN baselines on HCPYA and NKI-Rockland datasets, which verifies that the SFC is shaped and guided by the complex topological organization of the brain network.

The proposed HiM-SFC framework achieves the best overall performance across the three datasets, with 2.77 MAE on HCPYA and 12.86 MAE on NKI-Rockland, improving by 0.51 and 0.39 points over the second-best methods, respectively. The superior performance of the proposed framework across three datasets verifies that the hierarchical modular organization better captures age-related coupling, thereby improving the effectiveness and generalization of the multimodal framework.

\begin{table}[!t]
\centering
\setlength{\tabcolsep}{3pt}
\fontsize{8}{10}\selectfont
\renewcommand{\arraystretch}{0.9}
\caption{Ablation study on PMPool, AHCM, and CgC-Loss on HCPYA (MMSE prediction) and ADNI (disease classification).}
\label{Tab:Ablation—_modules}
\begin{tabular*}{\columnwidth}{@{\extracolsep{\fill}}lccc}
\toprule
% ===================== HCPYA block =====================
\multicolumn{4}{l}{\textbf{HCPYA: MMSE prediction}}\\
\midrule
\textbf{Method} & \textbf{MAE~$\downarrow$} & \textbf{RMSE~$\downarrow$} & \textbf{PCC~$\uparrow$}\\
\midrule
\textit{w/o } PMPool  & 2.02 (0.33) & 2.32 (0.41) & 0.12 (0.04)\\
\textit{w/o } AHCM    & 2.63 (0.31) & 3.35 (0.41) & 0.09 (0.03)\\
\textit{w/o } CgC-Loss  & 1.17 (0.18) & 1.57 (0.44) & 0.16 (0.04)\\
\midrule
\textbf{HiM-SFC(ours)}         & \textbf{1.01 (0.24)} & \textbf{1.25 (0.25)} & \textbf{0.25 (0.09)}\\
\midrule
% ===================== ADNI block =====================
\multicolumn{4}{l}{\textbf{ADNI: disease classification}}\\
\midrule
\textbf{Method} & \textbf{ACC~$\uparrow$} & \textbf{Sen~$\uparrow$} & \textbf{Spec~$\uparrow$} \\
\midrule
\textit{w/o } PMPool    & 0.72 (0.08) & 0.63 (0.11) & 0.73 (0.14) \\
\textit{w/o } AHCM      & 0.70 (0.09) & 0.56 (0.12) & 0.71 (0.15) \\
\textit{w/o } CgC-Loss  & 0.76 (0.07) & 0.67 (0.09) & 0.77 (0.10) \\
\midrule
\textbf{HiM-SFC(ours)}    & \textbf{0.78 (0.07)} & \textbf{0.69 (0.10)} & \textbf{0.81 (0.13)} \\
\bottomrule
\end{tabular*}
\end{table}

\begin{table}[!t]
\centering
\setlength{\tabcolsep}{3pt}
\fontsize{8}{10}\selectfont
\renewcommand{\arraystretch}{0.9}
\caption{Results of MMSE score prediction on HCPYA and disease classification on ADNI using different GNN encoders.}
\label{Tab:Ablation_GNN_encoders}
\begin{tabular*}{\columnwidth}{@{\extracolsep{\fill}}lccc}
\toprule
% ===================== HCPYA block =====================
\multicolumn{4}{l}{\textbf{HCPYA: MMSE prediction}}\\
\midrule
\textbf{Method} & \textbf{MAE~$\downarrow$} & \textbf{RMSE~$\downarrow$} & \textbf{PCC~$\uparrow$}\\
\midrule
GIN  & 1.11 (0.04) & 1.33 (0.17) & 0.18 (0.06)\\
GraphSAGE  & 1.05 (0.02) & 1.27 (0.15) & 0.20 (0.12)\\
%\midrule
\textbf{GCN}  & \textbf{1.01 (0.24)} &  \textbf{1.25 (0.25)} & \textbf{0.25 (0.09)}\\
\midrule
% ===================== ADNI block =====================
\multicolumn{4}{l}{\textbf{ADNI: disease classification}}\\
\midrule
\textbf{Method} & \textbf{ACC~$\uparrow$} & \textbf{Sen~$\uparrow$} & \textbf{Spec~$\uparrow$} \\
\midrule
GIN  & 0.75 (0.03) & 0.64 (0.13) & 0.78 (0.11) \\
GraphSAGE    & 0.77 (0.06) & 0.67 (0.13) & 0.79 (0.15) \\
%\midrule
\textbf{GCN}   & \textbf{0.78 (0.07)} & \textbf{0.69 (0.10)} & \textbf{0.81 (0.13)} \\
\bottomrule
\end{tabular*}
\end{table}

\begin{figure*}[!t]
    \centering
    \includegraphics[width=0.75\linewidth]{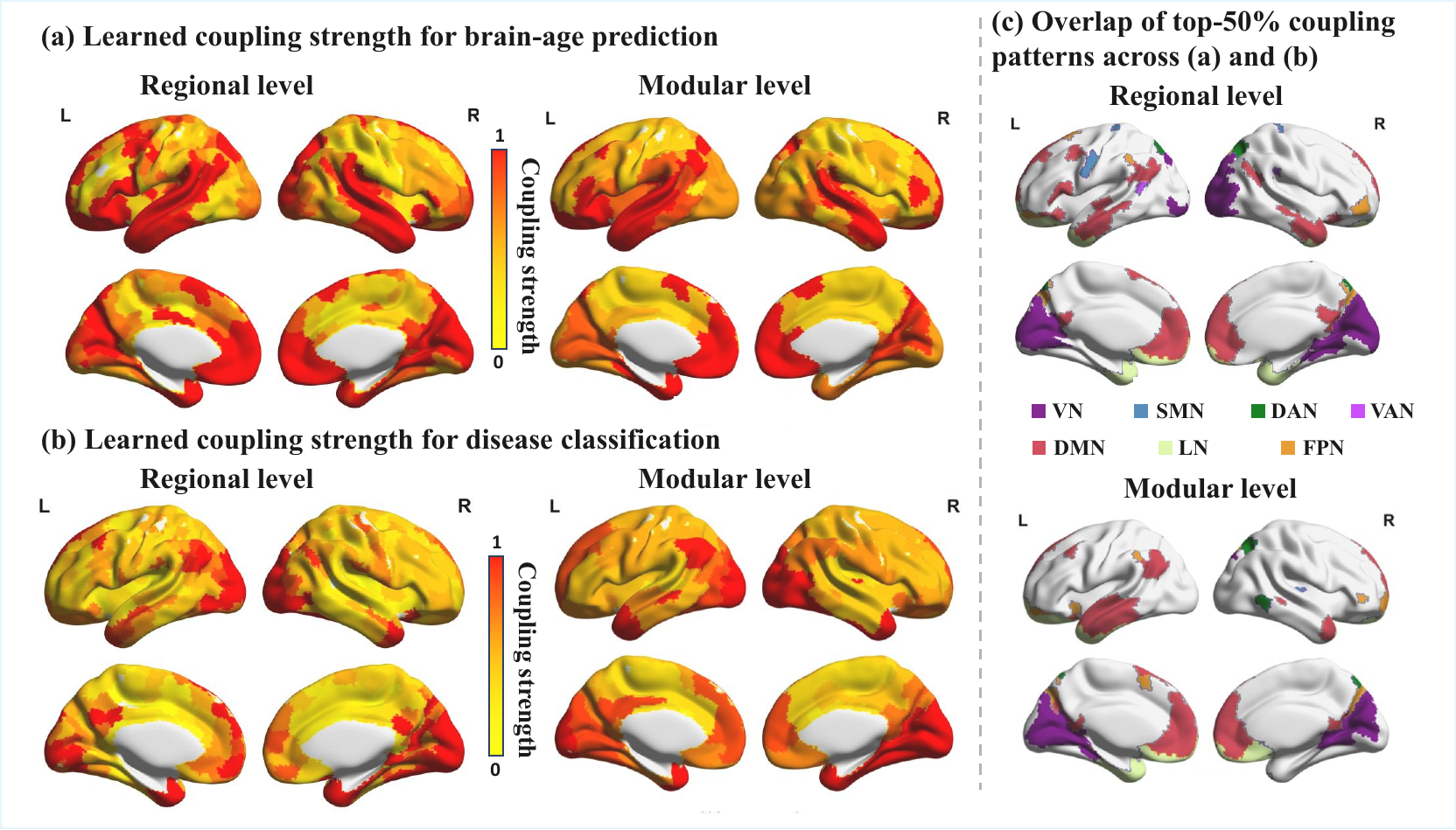}
    \caption{Multiscale coupling strength learned by the proposed model.
Regional-level maps show global coupling strength, whereas modular-level maps show local coupling strength, for (a) brain-age prediction in UKB and (b) disease classification in ADNI; (c) Overlapping of the top-50\% coupling pattern between brain age and disease prediction tasks at regional and modular levels, mapped to canonical seven modules (Visual (VN), Somatomotor (SMN), Dorsal Attention (DAN), Ventral Attention (VAN), Limbic (LN), Frontoparietal (FPN), and Default Mode (DMN) networks).}
    \label{fig:Coupling_ROIs_visualization}
\end{figure*}

\subsubsection{Cognitive score prediction}
%\noindent\textbf{Cognitive score prediction.} 
As seen from Table.~\ref{Tab:CompareCognitive}, HiM-SFC achieves consistently superior performance on the two datasets over the competing methods. In contrast to brain age prediction, methods that capture the coupling from a flat graph perspective, i.e., Joint-GCN and Cross-GNN, perform worse than GNN baselines on HCPYA, probably because they ignore the complex topology of the brain network and fail to learn coupling aligned with functional modules. In contrast, advanced coupling methods, e.g., IMG-GCN and CBGT, achieve better performance on cognitive score prediction. Specifically, IMG-GCN achieves 1.72 MAE on HCPYA and 1.84 MAE on UKB, respectively, probably because it incorporates functional modular organization to guide the learning of coupling patterns linked to cognition. The proposed HiM-SFC attains a mean absolute error of 1.01 on HCPYA and 1.66 on UKB, respectively, outperforming the second-best model by 0.47 and 0.25 in prediction error, demonstrating the benefit of modeling hierarchical modular coupling.

\subsubsection{Disease classification}
%\noindent\textbf{Cognitive score prediction.} 
The proposed method is evaluated on the ADNI dataset for disease classification. The results are presented in Table.~\ref{Tab:CompareDisease}. Our results indicate that coupling-based fusion methods, such as Joint-GCN, IMG-GCN, and CBGT, outperform the multimodal GNN baselines. Meanwhile, in addition to sensitivity, the proposed method consistently outperforms all competing methods. Specifically, the proposed method achieves an accuracy of 0.78 and a specificity of 0.81, indicating that the proposed hierarchical modular coupling framework is promising for capturing disease-related patterns.

%\FloatBarrier

\subsection{Ablation studies}

 %We further conduct ablation studies to quantify the contribution of each component in our framework and to examine sensitivity to key hyperparameters.

\label{section_ablation} 
\subsubsection{Key hyperparameters}
%\noindent\textbf{Ablation on modules.} 
To assess the influence of key hyperparameters on prediction performance, we conduct hyperparameter analyses on HCPYA and ADNI for MMSE score prediction and disease classification, respectively. The key hyperparameters include the pooling ratio of the PMPool, the number of PMPool hierarchies within the framework, and the regularization weight of the CgC-Loss. Fig.~\ref{fig:hyperparameter}a presents the results on HCPYA using MAE and RMSE, while Fig.~\ref{fig:hyperparameter}b presents the results on ADNI using accuracy (ACC), specificity (Spec), and sensitivity (Sen).

\noindent\textbf{The number of pooling ratios.} As introduced in Section~\ref{sec_PMPool}, the size of the assignment matrix \(S\in \mathbb{R}^{N\times k}\) is controlled by the number of prototypes \(k\), which guides the model to learn hierarchical modular organization at different resolutions. In this study, we use a pooling ratio \(r\) to determine \(k\) through \(k_l = N_{l-1} \times r_l\), where \(N_{l-1}\) is the number of nodes in layer \(l-1\). With this setting, a larger \(r_l\) leads to a larger \(k_l\), allowing PMPool to retain more modular communities. As shown in Fig.~\ref{fig:hyperparameter}, varying the pooling ratio can affect performance on both datasets. On HCPYA, the model starts from the flat baseline without PMPool (\(r=0\)) and achieves consistently better performance once hierarchical pooling is introduced, with the best MAE and RMSE obtained at \(r=0.12\). A similar trend is observed on ADNI, where ACC, Spec, and Sen all improve from the baseline and reach their best at \(r=0.12\).

\begin{table}[!t]
\centering
\setlength{\tabcolsep}{3pt}
\fontsize{8}{10}\selectfont
\renewcommand{\arraystretch}{0.9}
\caption{Results of MMSE score prediction on HCPYA and disease classification on ADNI using entmax and softmax.}
\label{Tab:Ablation_entmax}
\begin{tabular*}{\columnwidth}{@{\extracolsep{\fill}}lccc}
\toprule
% ===================== HCPYA block =====================
\multicolumn{4}{l}{\textbf{HCPYA: MMSE prediction}}\\
\midrule
\textbf{Method} & \textbf{MAE~$\downarrow$} & \textbf{RMSE~$\downarrow$} & \textbf{PCC~$\uparrow$}\\
\midrule
softmax  & 1.17 (0.03) & 1.33 (0.16) & 0.18 (0.04)\\
\textbf{entmax} & \textbf{1.01 (0.24)} &  \textbf{1.25 (0.25)} & \textbf{0.25 (0.09)}\\
\midrule
% ===================== ADNI block =====================
\multicolumn{4}{l}{\textbf{ADNI: disease classification}}\\
\midrule
\textbf{Method} & \textbf{ACC~$\uparrow$} & \textbf{Sen~$\uparrow$} & \textbf{Spec~$\uparrow$} \\
\midrule
softmax  & 0.76 (0.17) & 0.66 (0.12) & 0.77 (0.14) \\
\textbf{entmax} & \textbf{0.78 (0.07)} & \textbf{0.69 (0.10)} & \textbf{0.81 (0.13)} \\
\bottomrule
\end{tabular*}
\end{table}

% \begin{table}[!t]
% \centering
% \setlength{\tabcolsep}{3pt}
% \fontsize{8}{10}\selectfont
% \renewcommand{\arraystretch}{0.9}
% \caption{Results of MMSE score prediction on HCPYA and disease prediction on ADNI using entmax and softmax.}
% \label{Tab:Ablation_entmax}
% \begin{tabular*}{\columnwidth}{@{\extracolsep{\fill}}lccc}
% \toprule
% % ===================== HCPYA block =====================
% \multicolumn{4}{l}{\textbf{HCPYA: MMSE prediction}}\\
% \midrule
% \textbf{Method} & \textbf{MAE~$\downarrow$} & \textbf{RMSE~$\downarrow$} & \textbf{PCC~$\uparrow$}\\
% \midrule
% softmax  & 1.17 (0.03) & 1.33 (0.16) & 0.18 (0.04)\\
% \textbf{entmax} & \textbf{1.01 (0.24)} &  \textbf{1.25 (0.25)} & \textbf{0.25 (0.09)}\\
% \midrule
% % ===================== ADNI block =====================
% \multicolumn{4}{l}{\textbf{ADNI: disease prediction}}\\
% \midrule
% \textbf{Method} & \textbf{ACC~$\uparrow$} & \textbf{Sen~$\uparrow$} & \textbf{Spec~$\uparrow$} \\
% \midrule
% softmax  & 74.3 (0.17) & 51.8 (0.12) & 75.8 (0.14) \\
% \textbf{entmax} & \textbf{78.0 (0.07)} & \textbf{68.7 (0.10)} & \textbf{81.5 (0.13)} \\
% \bottomrule
% \end{tabular*}
% \end{table}

\noindent\textbf{The number of PMPool hierarchies.} We investigate how the PMPool hierarchies influence prediction performance. As illustrated in Fig.~\ref{fig:hyperparameter}, introducing hierarchical PMPool improves performance on both HCPYA and ADNI compared with the baseline setting without PMPool. On HCPYA, the best MMSE prediction is achieved with a single PMPool hierarchy, while deeper hierarchies gradually degrade performance. In ADNI, the best classification performance is also achieved at one hierarchy, with higher ACC, Spec, and Sen than in the baseline and deeper hierarchies. When the number of hierarchies increases to 2 and 3, performance declines on both datasets, which is probably due to over-smoothing caused by excessively stacked GNN layers \citep {li2018deeper}.

\noindent\textbf{The regularization weight of CgC-Loss.} In addition, we evaluate how the coupling-guided clustering loss \(L_{CC}\) affects performance under different regularization weights in the range of \([0.05:0.03:0.17]\). Here, the baseline framework includes only the PMPool and AHCM. As shown in Fig.~\ref{fig:hyperparameter}, incorporating the CgC-Loss improves performance on both datasets, and the best results are achieved when the regularization weight is set to 0.11. Specifically, on HCPYA, the lowest MAE and RMSE are obtained at 0.11, while on ADNI, ACC, Spec, and Sen also reach their best overall values at the same weight. %It is also worth noting that, even though the performance is worse than the baseline at 0.17, HiM-SFC still outperforms the best competing methods with 1.20 MAE and 1.43 RMSE, demonstrating the effectiveness in modelling the hierarchical modular coupling pattern.

\begin{figure*}[!t]
    \centering
    \includegraphics[width=0.7\linewidth]{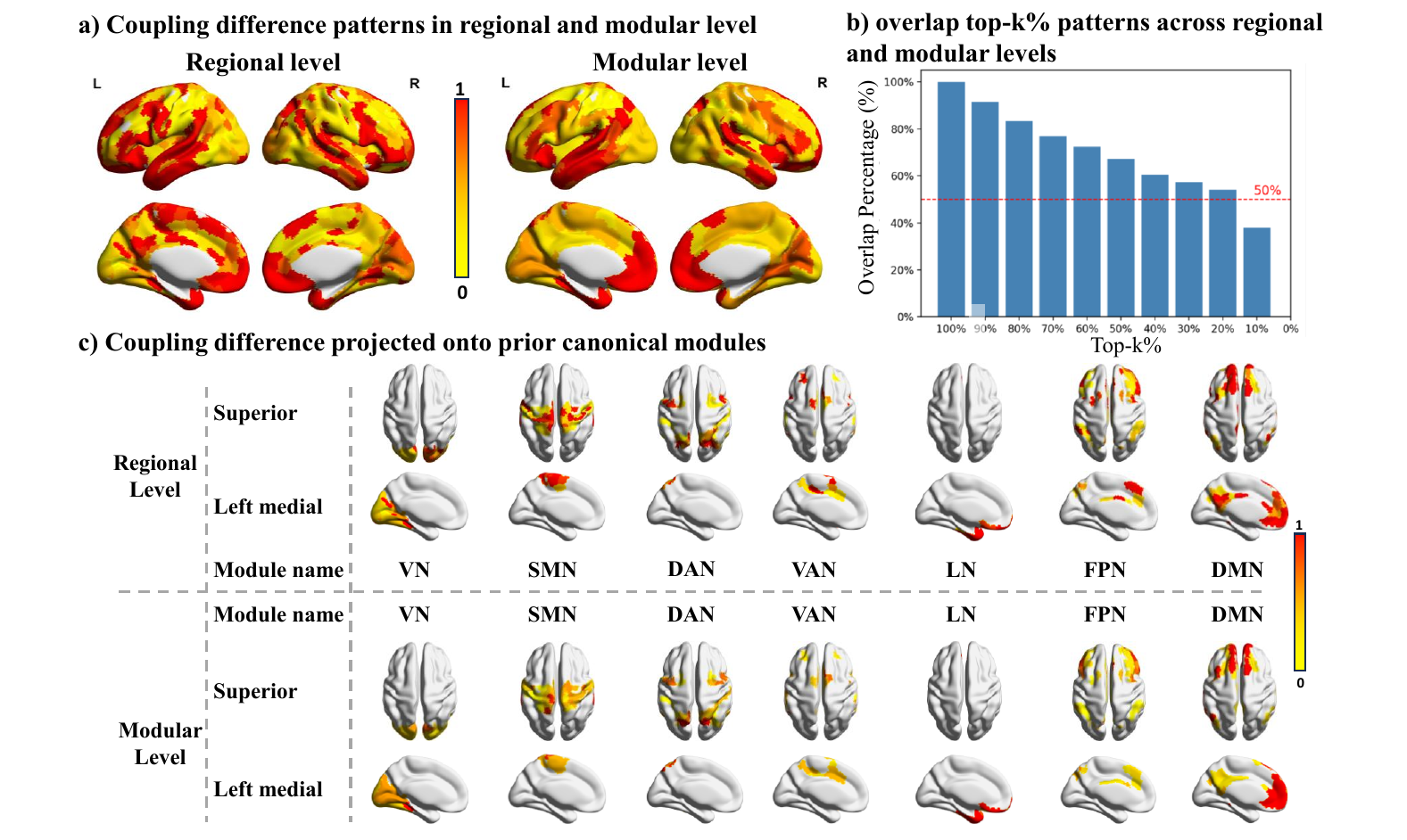}
    \caption{Visualization of age-group (40–50 vs. 60–70 years) differences in learned coupling strength in UKB at the regional (Schaefer 500 ROIs) and modular levels (a), the overlap percentage between regional and modular coupling-difference patterns across top-k\% ROIs (b), and their projection onto the canonical seven modules (Visual (VN), Somatomotor (SMN), Dorsal Attention (DAN), Ventral Attention (VAN), Limbic (LN), Frontoparietal (FPN), and Default Mode (DMN) networks) (c).}
    \label{fig:UKB_modules_visualization}
\end{figure*}

\subsubsection{Key components}
%\noindent\textbf{Ablation on key parameters.} 
To verify the efficacy of (1) PMPool, (2) AHCM, and (3) CgC-Loss, we conduct ablation studies on the HCPYA and ADNI datasets, additionally reporting PCC as a complementary evaluation metric. Results from Table.~\ref{Tab:Ablation—_modules} suggest that all three components contribute meaningfully. In HCPYA MMSE score prediction, removing AHCM results in the largest degradation, with 2.63 MAE and 3.35 RMSE, indicating that modeling hierarchical coupling is essential for integrating structural and functional brain connectomes. Ablating PMPool also degrades the performance, implying that the proposed prototype-based pooling can help learn the subject-level hierarchical modular organization. Removing CgC-Loss yields a smaller but still functional model, because it mainly serves as a regularization term to help learn the hierarchical modular structure driven by coupling. 

A similar trend is observed in ADNI disease classification. The largest decline occurs during AHCM ablation, highlighting the importance of coupling-aware interactions for connectome integration. The CgC-Loss brings a modest yet consistent decline across ACC, Sensitivity, and Specificity. Overall, the full HiM-SFC achieves the best and most balanced performance. 
% The ablation results on HCPYA MMSE show that all three components contribute to the model performance, and removing any of them degrades performance. Without PMPool, MAE rises from 1.01 to 2.02, and PCC drops from 0.21 to 0.12, suggesting that prototype-based modular pooling is important for extracting stable, informative multiscale modules. Dropping AHCM causes the greatest deterioration, with MAE increasing to 2.63 and PCC falling to 0.09, supporting the view that explicitly modelling hierarchical coupling is the main driver of predictive signal. Removing CgC-Loss also degrades performance, though less severely than removing AHCM, increasing MAE to 1.17 and reducing PCC to 0.16, indicating that coupling-guided clustering helps align the learned hierarchies across modalities and improves robustness.

\subsubsection{GNN encoders}
We also evaluate alternative GNN encoders within the proposed framework. As seen from the Table.~\ref{Tab:Ablation_GNN_encoders}, different GNN encoders may affect the performance, but do not diminish the overall advantage of HiM-SFC. In HCPYA MMSE score prediction, GCN achieves the best performance with an MAE of 1.01, an RMSE of 1.25, and a PCC of 0.25, outperforming GraphSAGE and GIN. A similar trend is observed on ADNI disease classification, where GCN achieves the best ACC with 0.78. Notably, the improvement over GraphSAGE is modest but consistent across both datasets, implying that the most contribution arises from the proposed coupling framework, while the GNN encoder primarily modulates how structural and functional signals are propagated. Overall, we use GCN as the default backbone for our framework, as it provides robust generalization across both regression and classification settings.

\subsubsection{Modular sparsity}
To evaluate the effectiveness of the sparse normalization operator, entmax, used in both PMPool and AHCM, we conduct the experiments on HCPYA and ADNI datasets. As shown in Table.~\ref{Tab:Ablation_entmax}, entmax yields a consistent improvement on both tasks. On the HCPYA MMSE prediction, entmax reduces MAE from 1.17 to 1.01 and RMSE from 1.33 to 1.25, whilst increasing PCC from 0.18 to 0.25. A similar pattern can be observed on the ADNI disease classification with a higher ACC. This advantage is likely attributed to the sparsity-inducing behavior of entmax, which can produce sparser assignments and attention maps~\citep{zhang2024constructing}. This aligns well with prior knowledge, as brain community structure is typically modular, with dense intra-community and sparse inter-community connectivity, while SFC is regionally heterogeneous rather than uniformly strong across the cortex~\citep{zhang2024constructing, zamani2022local}. In practice, sparse normalization, such as entmax, is likely to reduce spurious community assignments in PMPool and suppress noisy couplings in AHCM, thereby improving multimodal representations. Therefore, we adopt entmax as the default normalization throughout the framework.

% \begin{figure*}[!t]
%     \centering
%     \includegraphics[width=1\linewidth]{pdf/HCP_inter.pdf}
%     \caption{Visualization of learned structural and functional modules by PMPool from two different MMSE cognitive score groups (28 and 30) on HCPYA dataset.}
%     \label{fig:HCP_MMSE_modules}
% \end{figure*}

\begin{figure*}[!t]
    \centering
    \includegraphics[width=0.7\linewidth]{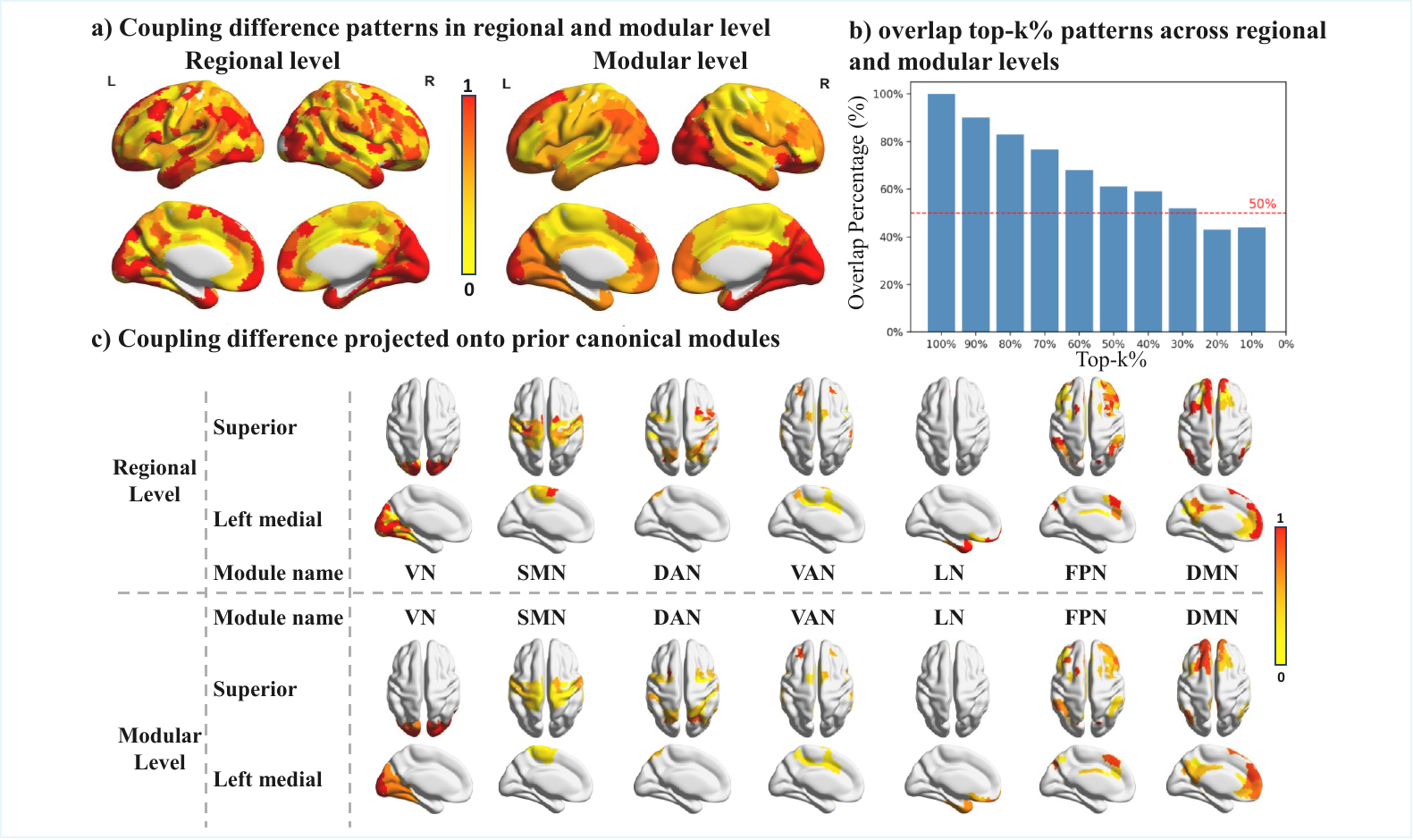}
    \caption{Visualization of subgroup differences (HC vs. dementia) in learned coupling strength in ADNI at the regional (Schaefer 500 ROIs) and modular levels (a), the overlap percentage between regional and modular coupling-difference patterns across top-k\% ROIs (b), and their projection onto the canonical seven modules (Visual (VN), Somatomotor (SMN), Dorsal Attention (DAN), Ventral Attention (VAN), Limbic (LN), Frontoparietal (FPN), and Default Mode (DMN) networks) (c).}
    \label{fig:ADNI_modules_visualzation}
\end{figure*}

\subsection{Interpretability}
\subsubsection{Learned hierarchical multiscale SFC}
\label{Learned SFC}

To examine the task-specific hierarchical multiscale coupling learned by AHCM, we estimate group-wise coupling strength at both regional and modular levels. Results are computed within each fold and averaged across the 5-fold training and testing splits for UKB and ADNI. At the regional level, we average the attention matrices produced by AHCM across subjects. For modular-level coupling, we derive group-wise communities from the structural and functional assignment matrices and average them across subjects, and apply spectral clustering to identify representative communities. These communities are then used to aggregate ROI-wise coupling strength into module-wise strength for visualization using BrainNet Viewer~\citep{di2023connectomes}. The resulting multiscale coupling patterns are shown in Fig.~\ref{fig:Coupling_ROIs_visualization}.

At the regional level, coupling strength is strongest in distributed association cortices, indicating that the learned interactions reflect widespread coordination. In brain-age prediction, stronger coupling is concentrated in medial and lateral prefrontal–parietal regions and posterior midline areas, consistent with circuits implicated in age-related reconfiguration of higher-order cognition~\citep{sala2015reorganization}. In contrast, the ADNI disease prediction shows stronger involvement of posterior and temporoparietal regions, indicating that AHCM emphasizes coupling signals that align with disease-sensitive systems~\citep{baum2020development}. At the modular level, these regional signals organize into clearer mesoscale patterns. After aggregating coupling through the learned community hierarchy and projecting onto prior functional modules, the maps become smoother and more spatially contiguous. This suggests that the model preserves regional variability while organizing coupling differences into clearer module-level patterns.

We further examine the cross-task consistency of the learned coupling patterns, shown in Fig.~\ref{fig:Coupling_ROIs_visualization}c. We identified the top 50\% ROIs with the strongest coupling strength for each task at both regional and modular levels. Overlapping ROIs were extracted and mapped onto the canonical seven modules~\citep{yeo2011organization} for neurobiological interpretation. Consistent overlaps were observed across both scales, primarily involving the visual cortex and default-mode network (DMN), which are associated with aging- and dementia-related brain changes. Alterations in visual network organization and reduced SFC in occipital regions have been reported in both normal aging and neurodegenerative conditions~\citep{andrews2007disruption}. The DMN, including the medial prefrontal and posterior cingulate areas, is known to be particularly vulnerable in Alzheimer’s disease and age-related cognitive decline~\citep{damoiseaux2006consistent, buckner2008brain}. These overlapping patterns suggest that the model captures biologically meaningful structure–function interactions shared across tasks.

\subsubsection{Difference of hierarchical multiscale coupling}

To examine differences in hierarchical multiscale coupling and modular organization, we conducted subgroup analyses. In UKB, participants were divided into two age subgroups (40–50 yrs: n=900; 60–70 yrs: n=1,058). In ADNI, the analysis was performed for healthy controls (n=72) and dementia patients (n=67). Following the procedure in Subsection~\ref{Learned SFC}, we computed subgroup-averaged coupling strength at both regional and modular levels, then calculated the differences between subgroups. The resulting difference maps were rescaled using min–max normalization for visualization. To assess whether the model captures consistent multiscale patterns, we quantified the spatial overlap across scales by selecting the top-k\% ROIs with the largest coupling differences and calculating their overlap across the range of [100\% : 10\% : 10\%]. For neurobiological interpretation, the coupling difference maps were projected onto the canonical seven modules~\citep{yeo2011organization}. The results are shown in Figs.~\ref{fig:UKB_modules_visualization} and ~\ref{fig:ADNI_modules_visualzation}.

As shown in Fig.~\ref{fig:UKB_modules_visualization}a, similar coupling difference patterns are observed between the two age groups at both regional and modular levels, particularly in the medial frontal cortex, lateral temporal areas, and posterior cortical regions. The consistency across scales suggests that the model captures coherent multiscale alterations in SFC. In Fig.~\ref{fig:UKB_modules_visualization}b, we quantify this similarity, showing that the overlap remains above 50\% even when considering only the top 20\% ROIs (100 regions), indicating that similar patterns are preserved at both scales. For interpretation, we project the coupling differences onto the canonical seven modules. As shown in Fig.~\ref{fig:UKB_modules_visualization}c, both representations exhibit similar coupling differences across the seven modules, with notable differences in the limbic, default-mode, and dorsal attention networks. These findings align with previous studies showing that aging preferentially affects higher-order association networks, particularly the default-mode and limbic systems, which are related to substantial structural and functional reorganization throughout the adult lifespan. Additionally, coupling modulation in control networks highlights the interconnectedness of functional systems in age-related changes~\citep{jockwitz2021resting,huang2022aging,wang2024aging,khalilian2024age}.

As shown in Fig.~\ref{fig:ADNI_modules_visualzation}a, similar coupling difference patterns are observed between healthy controls (HC) and dementia patients at both levels. Notably, coupling differences are seen in several cortical regions, particularly the medial frontal cortex, posterior medial areas, and lateral temporal regions, suggesting that the model captures coherent multiscale alterations in SFC in dementia. In Fig.~\ref{fig:ADNI_modules_visualzation}b, the overlap between regional and modular coupling-difference maps remains high across thresholds, with overlap exceeding 50\% even for the top 30\% ROIs (150 regions), indicating that similar patterns are preserved at both scales. For interpretation, we project the coupling differences onto the canonical seven modules. As shown in Fig.~\ref{fig:ADNI_modules_visualzation}c, both regional and modular representations exhibit consistent patterns across the seven modules, with notable effects in the visual, frontoparietal, and default-mode networks. These observations are consistent with prior studies indicating that neurodegenerative disorders, particularly Alzheimer’s disease, preferentially affect large-scale brain systems such as the default-mode, frontoparietal, and visual networks, which show substantial structural and functional reorganization along the disease continuum. Additionally, the modulation of coupling in control networks further reflects the interconnected nature of functional systems in neurodegeneration.~\citep{seeley2009neurodegenerative,franzmeier2019functional,jones2011age}.

\section{Conclusion}
In this work, we introduce a hierarchical multiscale structure–function coupling framework for multimodal brain connectome integration, motivated by the prior that SC–FC relationships are nonlinear and organized over nested modular hierarchies rather than a single flat graph. Building on this view, our method jointly learns (i) individualized, modality-specific multiscale community structures via Prototype-based Modular Pooling (PMPool), (ii) within- and cross-hierarchy structure–function interactions via an Attention-based Hierarchical Coupling Module (AHCM), and (iii) coupling-guided coordination of structural and functional hierarchies through a Coupling-guided Clustering loss (CgC-Loss), allowing cross-modal interactions to shape community structure. Across multiple cohorts and tasks, including age prediction, cognitive score prediction, and dementia classification, the proposed model consistently outperforms multimodal GNN baselines as well as recent coupling-based and hierarchy-aware methods. Ablation and sensitivity analyses further confirm the contributions of each component and identify stable operating regimes. Beyond performance, our interpretability analyses indicate that the learned coupling patterns and community assignments capture biologically plausible, group-specific signatures: both aging and dementia are associated with systematic shifts in high-coupling regions and reconfiguration of large-scale modules. Taken together, these results suggest that explicitly modeling hierarchical multiscale coupling is an effective approach for multimodal connectome fusion, improving generalization and interpretability while supporting studies of development, aging, and brain disorders. Future work will extend this framework to dynamically model hierarchical multiscale coupling and explore its potential to uncover mechanisms underlying a wider range of brain disorders.

\section{Acknowledgments}

The ADNI dataset used in this study were obtained from the Alzheimer’s Disease Neuroimaging Initiative (ADNI) database (adni.loni.usc.edu). The investigators within ADNI contributed to the design and implementation of ADNI and/or provided data, but did not participate in analysis or writing of this report. A complete list of ADNI investigators can be found at: \href{http://adni.loni.usc.edu/wp-content/uploads/how_to_apply/ADNI_Acknowledgement_List.pdf}. Data collection and sharing for ADNI is funded by the National Institute on Aging (National Institutes of Health Grant U19AG024904). This research has been conducted using the UK Biobank Resource under Application Number 52802. The authors are grateful to UK Biobank and all study participants for making the data available.

Chao Li acknowledge Guarantors of Brain. Data were provided [in part] by the Human Connectome Project, WU-Minn Consortium (Principal Investigators: David Van Essen and Kamil Ugurbil; 1U54MH091657) funded by the 16 NIH Institutes and Centers that support the NIH Blueprint for Neuroscience Research; and by the McDonnell Center for Systems Neuroscience at Washington University.
% Acknowledgments for this work...

%%Harvard
%\clearpage 
\bibliographystyle{model2-names.bst}\biboptions{authoryear}
\bibliography{refs}

% \section*{Supplementary Material}

\end{document}